\documentstyle[amsmath,a4,11pt]{article}

\date{}

\begin{document}

\title{{\bf Gravitational waves and cosmic magnetism; a
cosmological approach}}

\author{Christos G. Tsagas\thanks{e-mail address:
ctsagas@maths.uct.ac.za}\\{\small Department of Mathematics and
Applied Mathematics, University of Cape Town}\\ {\small Rondebosch
7701, South Africa}}

\maketitle

\begin{abstract}
We present the formalism for the covariant treatment of
gravitational radiation in a magnetized environment and discuss
the implications of the field for gravity waves in the
cosmological context. Our geometrical approach brings to the fore
the tension properties of the magnetic force lines and reveals
their intricate interconnection to the spatial geometry of a
magnetised spacetime. We show how the generic anisotropy of the
field can act as a source of gravitational wave perturbations and
how, depending on the spatial curvature distortion, the magnetic
tension can boost or suppress waves passing through a magnetized
region.
\end{abstract}

\section{Introduction}
Magnetic fields appear everywhere in the universe. Planets, stars
and galaxies carry fields that are large and extensive. The Milky
Way has a magnetic field, coherent along the plane of the galaxy,
with strength of a few $\mu G$. Magnetic fields also permeate the
intracluster region of galaxy clusters, extending well beyond the
core region of the cluster. In addition, reports of Faraday
rotation in high redshift Lyman-$\alpha$ systems suggest that
dynamically significant magnetic fields might have been present at
high redshift condensations. In short, the more we look for
magnetic fields in the universe the more ubiquitous we find them
to be~\cite{K}-\cite{GR}. Gravitational waves, on the other hand,
are as elusive as ever. An inevitable prediction of general
relativity, gravitational waves are propagating ripples in the
spacetime fabric triggered by changes in the matter distribution.
Their extremely weak coupling to matter, however, means that
detecting these ripples is a formidable task. Given the ubiquity
of magnetic fields in the universe and their strong presence near
some of the most promising candidates for detectable gravity wave
signals, understanding the interaction between the two sources
becomes especially interesting. Cosmology could provide the
grounds for an exploratory first step, and recently there were two
attempts in this direction~\cite{MTU,CD}. The former employed the
covariant formalism to investigate the evolution of gravitational
waves in a magnetised cosmological environment. The latter adopted
a metric based approach to study gravity wave production by
stochastic large-scale magnetic fields.\footnote{The issue of
magnetically induced tensor (i.e.~gravitational wave) signatures
in the Cosmic Microwave Background spectrum has been raised and
discussed in~\cite{DFK,MKK} and also in~\cite{MTU}.} Here,
following on the work of~\cite{MTU}, we attempt to revisit the
issue. Our geometrical approach brings to the fore the vector
nature and the tension properties of magnetic fields and reveals
the intricate interconnection between magnetic tension and spatial
geometry. A direct implication of this coupling is that the
magnetic contribution to spatial curvature deformations, along the
field's own direction, is always zero.

We begin with a brief outline of the covariant formalism and then
apply it to the treatment of gravitational wave perturbations in a
weakly magnetized, almost-FRW, cosmological environment. Our
background universe is a non-magnetized, spatially flat FRW
cosmology filled with a single perfect fluid of very high
conductivity. This model is perturbed by allowing for weak
gravitational waves and a weak magnetic field and the impact of
the field on the evolution of the waves is analysed. We introduce
no a priori relation between the wave and the magnetic
anisotropies, which enables us to look beyond the gravity-wave
production issue. Throughout the analysis we employ covariantly
defined tensors and use scalars that are invariantly constructed
from these tensors. This means that we can account for the full
spectrum of the directionally dependent magnetic effects,
particularly those coming from the tension properties of the field
lines. Confining to large scales we calculate the magnetic effects
analytically both in the radiation and the dust dominated eras. We
find that the presence of the field leaves the linear evolution of
the waves unaffected, but modifies their magnitude. This implies
that the overall magnetic impact depends entirely on the initial
conditions. In the absence of gravity waves, we find that the
anisotropic nature of the field leads to wave production. In
general, however, the magnetic effect is particularly sensitive to
the initial curvature deformation as measured along the direction
of the field lines. This is where the subtle role of the magnetic
tension becomes important. A negative curvature perturbation is
found to boost the energy of the waves, whereas positive curvature
leads to a damping effect. In either case the tension properties
of the field tend to keep the spatial curvature deformation down
to a minimum.

The layout of the paper is as follows. In Secs.~2 and 3 we provide
a brief description of the covariant treatment of cosmological
perturbations and of cosmic magnetic fields respectively. Section
4 presents the necessary formalism for the study of gravitational
waves within weakly magnetized almost-FRW cosmologies. The
equations for the linear evolution of the wave's energy density
are given in Sec.~5, and in Sec.~6 we look into the magnetic
effects on gravitational radiation during the radiation and the
dust dominated eras.

We employ a metric with signature $(-+++)$, the spacetime indices
take the values $0,1,2,3$ and use geometrised units with $c=1=8\pi
G$. The Riemann and Ricci tensors are fixed by
$2\nabla_{[a}\nabla_{b]}u_c=R_{abcd}u^d$ and $R_{ab}=R^c{}_{acb}$
respectively.

\section{Covariant description of cosmological perturbations}
The Ellis-Bruni approach to cosmological perturbations~\cite{EB},
follows earlier studies by Hawking~\cite{H} and Olson~\cite{O},
and is based on Ehlers' work on covariant hydrodynamics~\cite{E}.
For a detailed and updated presentation of the covariant formalism
the reader is referred to~\cite{EvE}. The essence of the method is
to identify a set of covariantly defined variables that describe
the inhomogeneity and the anisotropy of the universe in a
transparent and gauge invariant manner. The non-linear equations
governing the dynamics of these variables are then obtained from
the full field equations. The transparency of the covariant
variables means that the physical and geometrical content of their
dynamical equations is simple to extract. The non-linear formulae
can be linearised about a chosen background yielding a set of
equations that describe deviations from inhomogeneity and isotropy
in a straightforward way.

In the covariant approach all physical and geometrical quantities
are decomposed with respect to a fundamental timelike velocity
field $u_a$. The latter is determined by the motion of the matter
in the universe and introduces a unique time plus space (1+3)
threading of the spacetime, as opposed to the 3+1 ADM slicing.
Every observer has an instantaneous three-dimensional rest space,
which is the tangent hypersurface orthogonal to $u_a$. In a
general cosmological model $u_a$ is chosen so that it reduces to
the preferred velocity at the FRW limit, thus guaranteeing the
gauge invariance of the formalism. The metric
$h_{ab}=g_{ab}+u_au_b$ of the observer's rest space ($g_{ab}$ is
the spacetime metric) also defines the projected covariant
derivative ${\rm D}_a$ according to ${\rm
D}_aT_{b\ldots}{}^{c\ldots}=h_a{}^dh_b{}^e\ldots
h^c{}_f\ldots\nabla_dT_{e\ldots}{}^{f\ldots}$. When $u_a$ is
irrotational, ${\rm D}_a$ reduces to the covariant derivative in
the hypersurfaces orthogonal to the observer's world line. It is
also convenient to introduce the derivative
$\dot{T}_{b\ldots}{}^{c\ldots}=u^a\nabla_aT_{b\ldots}{}^{c\ldots}$
along the flow lines of $u_a$. With these definitions, the
covariant derivative of the fundamental velocity field decopmoses
as
\begin{equation}
\nabla_bu_a=\sigma_{ab}+ \omega_{ab}+ {\textstyle{1\over3}}\Theta
h_{ab}- \dot{u}_au_b\,. \label{Dbua}
\end{equation}
The above defines the shear $\sigma_{ab}={\rm D}_{\langle
b}u_{a\rangle}$, the vorticity $\omega_{ab}={\rm D}_{[b}u_{a]}$,
the (volume) expansion scalar $\Theta={\rm D}^au_a$ and the
4-acceleration $\dot{u}_a=u^b\nabla_bu_a$ associated with the
observer's motion.\footnote{Round brackets indicate symmetrization
and square ones antisymmetrisation. Angled brackets indicate the
projected, symmetric, trace-free part (PSTF) of second rank
tensors (i.e.~$S_{\langle
ab\rangle}=[h_{(a}{}^ch_{b)}{}^d-(1/3)h^{cd}h_{ab}]S_{ab}$) and
the projected part of vectors (i.e.~$V_{\langle
a\rangle}=h_a{}^bV_b$)~\cite{MGE}.} Introducing the projected
totally antisymmetric tensor $\varepsilon_{abc}=\eta_{abcd}u^d$,
where $\eta_{abcd}$ is the spacetime alternating tensor, we define
the vorticity vector $\omega_a=\varepsilon_{abc}\omega^{bc}/2$.
The latter is also written as $\omega_a=-{\rm curl}u_a/2$ with
${\rm curl}u_a=\varepsilon_{abc}{\rm D}^bu^c$. In addition we
introduce the generalised curl of tensors by ${\rm
curl}T_{ab\ldots c}=\varepsilon_{de(a}{\rm D}^dT_{b\ldots
c)}{}^e$. The volume expansion defines an average length scale
$a$, namely the scale factor of the universe, via
$\dot{a}/a=\Theta/3$. The variables $\sigma_{ab}$, $\omega_a$, and
$\dot{u}_a$ characterise anisotropy in the local expansion and
vanish identically in the FRW limit. The projected gradients of
scalars, vector and tensors describe local inhomogeneity in the
observer's instantaneous rest space and also vanish in exact FRW
spacetimes.

An additional key decomposition is that of the matter
stress-energy tensor $T_{ab}$. Relative to a fundamental observer,
\begin{equation}
T_{ab}=\mu u_au_b+ ph_{ab}+ 2q_{(a}u_{b)}+ \pi_{ab}\,,
\label{Tab}
\end{equation}
where $\mu=T_{ab}u^au^b$ is the energy density, $p=h_{ab}T^{ab}/3$
is the isotropic pressure, $q_a=h_a{}^bT_{bc}u^c$ is the energy
flux and $\pi_{ab}=T_{\langle ab\rangle}$ are the anisotropic
stresses of the matter component. In the FRW limit $q_a$ and
$\pi_{ab}$ vanish, as the energy-momentum tensor is always of the
perfect-fluid form.

\section{The magnetic field}
\subsection{Covariant description of cosmic magnetic fields}
The covariant description of electromagnetic fields was given
in~\cite{El}, and the Ellis-Bruni approach was applied to
magnetized cosmological perturbations in~\cite{TB1,TM1}.
Covariantly, the electromagnetic Faraday tensor ($F_{ab}$)
decomposes into an electric ($E_a$) and a magnetic ($H_a$) field
as
\begin{equation}
F_{ab}=u_{[a}E_{b]}+\varepsilon_{abc}H^c\,,  \label{Fab}
\end{equation}
implying that $E_a=F_{ab}u^a$ and $H_a=\varepsilon_{abc}F^{bc}/2$.
Assuming that the universe is described by an infinitely
conductive medium during most of its lifetime, we can ignore the
presence of the electric field. Then Maxwell's equations reduce to
one propagation equation
\begin{equation}
\dot{H}_{\langle a\rangle}=-{\textstyle{2\over3}}\Theta H_a+
\sigma_{ab}H^b+ \varepsilon_{abc}H^b\omega^c\,,  \label{M1}
\end{equation}
and three constraints
\begin{eqnarray}
\varepsilon_{abc}\dot{u}^bH^c+ {\rm curl}H_a&=&J_{\langle
a\rangle}\,,  \label{M2}\\ 2\omega_aH^a&=&\rho_{\rm e}\,,  \label{M3}\\
{\rm D}_aH^a&=&0\,,  \label{M4}
\end{eqnarray}
where $J_{\langle a\rangle}=h_a{}^bJ_b$ is the projected current
density and $\rho_{\rm e}=-u_aJ^a$ is the charge
density~\cite{TB1}. The above determine the evolution of the
magnetic field completely. Relative to the fundamental observer,
the stress-energy tensor of the field decomposes as
\begin{equation}
{\cal T}_{ab}={\textstyle{1\over2}}H^2u_au_b+
{\textstyle{1\over6}}H^2h_{ab}+ \Pi_{ab}\,,  \label{Tm}
\end{equation}
where $H^2=H_aH^a$ and
\begin{equation}
\Pi_{ab}=-H_{\langle a}H_{b\rangle}=
{\textstyle{1\over3}}H^2h_{ab}- H_aH_b\,.  \label{Pi}
\end{equation}
Note that in the absence of electric fields the electromagnetic
Poynting vector vanishes. Thus, the magnetic field behaves as a
special imperfect fluid with energy density $\mu_{\rm m}=H^2/2$,
isotropic pressure $p_{\rm m}=H^2/6$ and anisotropic pressure
$\Pi_{ab}$. The latter reflects the vector nature of the field and
carries the tension properties of the magnetic force lines.

\subsection{The magnetic tension}
Magnetic fields exert an isotropic pressure in all directions and
carry a tension along their lines of force, with each flux tube
behaving like an infinitely elastic rubber band~\cite{P,Me}. The
tension properties are encoded in the eigenvalues of $\Pi_{ab}$.
Orthogonal to $H_a$ there are two positive eigenvalues equal to
1/3 each. Thus, the magnetic pressure perpendicular to the field
lines is positive, reflecting their tendency to push each other
apart. In the $H_a$ direction, however, the eigenvalue is -2/3 and
the magnetic pressure is negative. The minus sign reflects the
elasticity of the magnetic lines and their tendency to remain as
``straight'' as possible. It should be emphasized that the total
magnetic pressure along the direction of the field lines is also
negative and equals $p_{\rm mg}=-\mu_{\rm mg}=-H^2/2$. In other
words, the false vacuum condition is satisfied along the magnetic
lines of force.

The magnetic tension has also non-trivial implications for the
geometry of a magnetised 3-dimensional space. Consider, for
example, the non-linear Gauss-Codacci equation associated with a
non-rotating magnetized spacetime. The Ricci curvature tensor of
the spatial hypersurfaces decomposes irreducibly to~\cite{TB1}
\begin{equation}
{\cal R}_{ab}={\textstyle{1\over3}}{\cal R}h_{ab}+
{\textstyle{1\over2}}\Pi_{ab}-
{\textstyle{1\over3}}\Theta\sigma_{ab}+ \sigma_{c\langle
a}\sigma^c{}_{b\rangle}+ E_{ab}\,,  \label{GC}
\end{equation}
where
\begin{equation}
{\cal R}=2\left(\mu+{\textstyle{1\over2}}H^2\right)
-{\textstyle{2\over3}}\Theta^2 +2\sigma^2\,,  \label{cR}
\end{equation}
is the associated Ricci scalar and we have assumed a infinitely
conductive perfect fluid for simplicity. Ignoring all sources but
the magnetic field and then contracting twice along the field
lines we obtain
\begin{equation}
\Re\equiv{\cal R}_{ab}\eta^a\eta^b=
{\textstyle{1\over3}}H^2+{\textstyle{1\over2}}\Pi_{ab}\eta^a\eta^b=
0\,, \label{cGC1}
\end{equation}
on using expression (\ref{Pi}) for $\Pi_{ab}$. Note that
$\eta_a=H_a/\sqrt{H^2}$ is the unit vector parallel to the
magnetic force lines (i.e.~$\eta_au^a=0$ and $\eta_a\eta^a=1$).
Hence, the tension properties of the field ensure that the spatial
curvature along the magnetic direction is unaffected by the energy
input of the field. This result demonstrates the generic tendency
of the field lines to remain straight and it is indicative of what
one might call a natural magnetic preference flat spatial
geometry. It is also independent of the energy density of the
field, namely of how close together or far apart the magnetic
lines are. Thus, no matter how much energy one pumps into or
removes from the field the magnetic contribution to ${\cal
R}_{ab}\eta^a\eta^b$ remains zero.

It should be emphasised that, despite their directional
dependence, the tension properties of the field can affect average
scalars such as the volume expansion of the universe
(see~\cite{MT}-\cite{T1}). As it turns out, they can also affect
the energy density of gravitational waves passing through a
magnetised region.

\section{Covariant description of magnetized gravitational waves}
\subsection{The background equations}
Consider an unperturbed non-magnetized FRW universe filled with a
single barotropic fluid of infinite conductivity. When the spatial
sections are flat, the background model is described by two
propagation equations
\begin{eqnarray}
\dot{\mu}&=&-(1+w)\Theta\mu\,,  \label{ce}\\
\dot{\Theta}&=&-{\textstyle{1\over3}}\Theta^2+
{\textstyle{1\over2}}\mu(1+3w)\,, \label{Ee}
\end{eqnarray}
and one constraint
\begin{equation}
\mu={\textstyle{1\over3}}\Theta^2\,,  \label{Fe}
\end{equation}
where $w=p/\mu$. Assuming that the fluid retains the barotropic
equation of state (i.e.~$p=p(\mu)$ always) and remains highly
conductive, we perturb the background allowing for weak
gravitational waves and a weak magnetic field. The infinite
conductivity of the medium means that we can disregard any
large-scale electric fields. The weakness of the magnetic field
allows us to treat its energy density, its isotropic pressure and
also the anisotropic magnetic stresses as first order
perturbations. This automatically ensures that all three of them
are gauge-independent variables. Finally, we assume that the
magnetic field is coherent on all scales of interest.

\subsection{The linear equations}
The covariant description of gravitational waves, in the absence
of magnetic fields, was originally considered by Hawking~\cite{H},
while more recent treatments can be found
in~\cite{DBE,C}.\footnote{Although the magnetic field can be
treated as a viscous fluid, the approach discussed in~\cite{H} is
not applicable here, since we are not introducing any
phenomenological relation between the shear and the anisotropic
stresses.} Covariantly, we monitor gravity waves via the electric
($E_{ab}=E_{\langle ab\rangle}$) and magnetic ($H_{ab}=H_{\langle
ab\rangle}$) components of the Weyl (or conformal curvature)
tensor $C_{abcd}$. The latter describes the locally free
gravitational field, namely tidal forces and gravity waves. For a
fundamental observer the Weyl tensor decomposes as~\cite{M2}
\begin{equation}
C_{abcd}=
\left(g_{abqp}g_{cdsr}-\eta_{abqp}\eta_{cdsr}\right)u^qu^sE^{pr}-
\left(\eta_{abqp}g_{cdsr}+g_{abqp}\eta_{cdsr}\right)u^qu^sH^{pr}\,,
\label{Weyl}
\end{equation}
where $g_{abcd}=g_{ac}g_{bd}-g_{ad}g_{bc}$ is the de Witt
supermetric and $\eta_{abcd}$ is the 4-dimensional permutation
tensor. It follows that
\begin{eqnarray}
E_{ab}&=&C_{acbd}u^cu^d\,,  \label{Eab}\\
H_{ab}&=&{\textstyle{1\over2}}\varepsilon_{acd}C_{be}{}{}^{cd}u^e\,,
\label{Hab}
\end{eqnarray}
where $\varepsilon_{abc}=\eta_{abcd}u^d$ is the alternating tensor
on the observer's 3-dimensional rest space. The electric part of
the Weyl tensor plays the role of the tidal tensor associated with
the Newtonian gravitational potential, while $H_{ab}$ is essential
for the propagation of gravitational radiation. Given that the
Weyl tensor vanishes in FRW spacetimes, $E_{ab}$ and $H_{ab}$
provide a covariant and gauge invariant description of
perturbations in the gravitational field. The electric and
magnetic components of $C_{abcd}$ also support the different
polarisation states of propagating gravitational radiation and
obey evolution equations remarkably similar to Maxwell's
formulae~\cite{MB}. In the presence of a weak magnetic field, the
linear evolution of $E_{ab}$ and $H_{ab}$ is determined by the
system
\begin{eqnarray}
\dot{E}_{ab}&=&-\Theta E_{ab}+
{\textstyle{1\over2}}\Theta\Pi_{ab}-
{\textstyle{1\over2}}\mu(1+w)\sigma_{ab}+ {\rm curl}H_{ab}\,
\label{lEab}\\ \dot{H}_{ab}&=&-\Theta H_{ab}- {\rm curl}E_{ab}+
{\textstyle{1\over2}}{\rm curl}\Pi_{ab}\,, \label{lHab}\\
\dot{\sigma}_{ab}&=&-{\textstyle{2\over3}}\Theta\sigma_{ab}-
E_{ab}+ {\textstyle{1\over2}}\Pi_{ab}+ {\rm D}_{\langle
a}\dot{u}_{b\rangle}\,, \label{lsigma}\\
\dot{\Pi}_{ab}&=&-{\textstyle{4\over3}}\Theta\Pi_{ab}\,,
\label{lPiab}
\end{eqnarray}
supplemented by the constrains
\begin{eqnarray}
{\rm D}^bE_{ab}&=&{\textstyle{1\over3}}{\rm D}_a\mu+
{\textstyle{1\over4}}{\rm D}_aH^2-
{\textstyle{1\over2}}\varepsilon_{abc}H^b{\rm curl}H^c\,,
\label{con1}\\ {\rm D}^bH_{ab}&=&\mu(1+w)\omega_a\,,
\label{con2}\\ {\rm D}^b\Pi_{ab}&=&\varepsilon_{abc}H^b{\rm
curl}H^c- {\textstyle{1\over6}}{\rm D}_aH^2\,,  \label{con3}\\
{\rm D}^b\sigma_{ab}&=&{\textstyle{2\over3}}{\rm D}_a\Theta+ {\rm
curl}\omega_a\,.  \label{con4}
\end{eqnarray}
Finally, one should keep in mind that to linear order
\begin{equation}
H_{ab}={\rm curl}\sigma_{ab}+ {\rm D}_{\langle
a}\dot{u}_{b\rangle}\,.  \label{con5}
\end{equation}

\subsection{Isolating the tensor perturbations}
In the absence of the magnetic field one isolates the pure tensor
perturbations of $E_{ab}$ and $H_{ab}$, namely the gravitational
waves, by demanding that
\begin{equation}
\omega_a=0={\rm D}_a\mu\,,  \label{tperts1}
\end{equation}
at all times. The above constraints set all the linear scalar and
vector perturbations to zero, while ensuring that the remaining
tensor fields are all transverse (i.e.~${\rm D}^aE_{ab}={\rm
D}^aH_{ab}={\rm D}^b\sigma_{ab}=0$). In the magnetic presence,
however, one needs to impose two additional constrains
\begin{equation}
{\rm D}_aH^2=0=\varepsilon_{abc}H^b{\rm curl}H^c\,.
\label{tperts2}
\end{equation}
In other words, the spatial gradients of the magnetic energy
density vanish and the field is also force free. Restrictions
(\ref{tperts2}) ensure that constraints (\ref{tperts1}) also hold
in the presence of the magnetic field (see Eqs.~(37), (41)
in~\cite{TB2}). Together, conditions (\ref{tperts1}),
(\ref{tperts2}) imply that ${\rm D}_ap=0$, $\dot{u}_a=0$ and ${\rm
D}_a\Theta=0$. In particular, the 3-gradients of the pressure
vanish as a result of the barotropic equation of state. The
vanishing of the 4-acceleration comes from the linearised
momentum-density conservation law (see Eq.~(87) in~\cite{TB1}).
Finally, the linear propagation equation of gradients in the
matter density guarantees that ${\rm D}_a\Theta=0$ to first order
(see Eqs.~(90), (91) in~\cite{TB1}). The same sets of constrains
also guarantee the transversality of all the associated tensor
fields, as Eqs.~(\ref{con1})-(\ref{con4}) immediately verify.

\section{Linear magnetized gravitational waves}
\subsection{Evolution of the wave anisotropy}
Having isolated the tensor modes, we drop the acceleration terms
from Eqs.~(\ref{lsigma}) and (\ref{con5}). Moreover, the spatial
flatness of the background means that the first order relation
$H_{ab}={\rm curl}\sigma_{ab}$ translates into ${\rm
curl}H_{ab}=-{\rm D}^2\sigma_{ab}$, where ${\rm D}^2={\rm D}_a{\rm
D}^a$ is the projected Laplacian operator. It follows that we can
eliminate the magnetic Weyl tensor from Eq.~(\ref{lEab}) in favour
of the shear and therefore reduce the total number of the
equations by one. Thus, the linear evolution of the magnetized
gravity waves is monitored by
\begin{eqnarray}
\dot{\sigma}_{ab}&=&-{\textstyle{2\over3}}\Theta\sigma_{ab}-
E_{ab}+ {\textstyle{1\over2}}\Pi_{ab}\,,
\label{lsab1}\\\dot{E}_{ab}&=&-\Theta E_{ab}+
{\textstyle{1\over2}}\Theta\Pi_{ab}-
{\textstyle{1\over2}}\mu(1+w)\sigma_{ab}- {\rm D}^2\sigma_{ab}\,,
\label{lEab1}\\
\dot{\Pi}_{ab}&=&-{\textstyle{4\over3}}\Theta\Pi_{ab}\,.
\label{lPiab1}
\end{eqnarray}
According to Eqs.~(\ref{lsab1}) and (\ref{lEab1}), the field acts
as a source of gravitational wave perturbations. Indeed, even when
$\sigma_{ab}=0=E_{ab}$ initially,
$\dot{\sigma}_{ab}\,,\dot{E}_{ab}\neq0$ because of the magnetic
presence. This is not surprising at all, given that magnetic
fields are natural sources of anisotropic stresses. As we shall
see later, however, the specific form of these stresses (i.e.~the
tension properties of the magnetic field lines) means that the
field presence can also suppress gravity wave distortions.

Confining to large scales we can ignore the Laplacian term in the
right-hand side of Eq.~(\ref{lEab1}) and reduce the above given
system to
\begin{eqnarray}
\dot{\sigma}_{ab}&=&-{\textstyle{2\over3}}\Theta\sigma_{ab}-
E_{ab}+ {\textstyle{1\over2}}\Pi_{ab}\,,
\label{lsab2}\\\dot{E}_{ab}&=&-\Theta E_{ab}+
{\textstyle{1\over2}}\Theta\Pi_{ab}-
{\textstyle{1\over2}}\mu(1+w)\sigma_{ab}\,, \label{lEab2}\\
\dot{\Pi}_{ab}&=&-{\textstyle{4\over3}}\Theta\Pi_{ab}\,.
\label{lPiab2}
\end{eqnarray}

\subsection{Evolution of the wave energy}
Consider the magnitudes $\sigma^2=\sigma_{ab}\sigma^{ab}/2$ and
$E^2=E_{ab}E^{ab}/2$ of the shear and the electric Weyl tensors
respectively. Once the pure tensor perturbations have been
isolated, these are the only scalars one can invariantly construct
from $\sigma_{ab}$ and $E_{ab}$. Moreover, $\sigma^2$ and $E^2$
provide a direct measure of the wave's energy density and
amplitude\footnote{The energy density of gravitational radiation
is determined by the pure tensor (i.e.~the traceless and
transverse - TT) part $H^{{_{\rm TT}}}_{\alpha\beta}$ of the
metric perturbation (e.g.~see~\cite{CD})
\begin{equation}
\rho_{_{\rm GW}}=
{\textstyle{1\over2}}\frac{(H_{\alpha\beta}^{_{\rm TT}})'(H_{_{\rm
TT}}^{\alpha\beta})'}{a^2}\,, \label{rhoGW}
\end{equation}
where $a$ is the dimensionless scale factor,
$\alpha\,,\beta=1,2,3$ and the dash indicates conformal time
derivatives. In a comoving frame the shear of the fluid flow is a
3-tensor (i.e.~$\sigma_{00}=0=\sigma_{0\alpha}$). Also, the pure
tensor part of the covariantly defined shear is related to the
traceless ransverse component of the metric perturbation by
$\sigma_{\alpha\beta}^{_{\rm TT}}=a(H_{\alpha\beta}^{_{\rm
TT}})'$, with $\sigma_{_{\rm TT}}^{\alpha\beta}=a^{-3}(H_{_{\rm
TT}}^{\alpha\beta})'$ (see~\cite{G,BDE}). On using these relations
Eq.~(\ref{rhoGW}) gives
\begin{equation}
\rho_{_{\rm GW}}= \sigma_{_{\rm TT}}^2\,, \label{crhoGW}
\end{equation}
with $\sigma_{_{_{\rm TT}}}^2=\sigma_{\alpha\beta}^{_{\rm
TT}}\sigma_{_{\rm TT}}^{\alpha\beta}/2$.} Their large-scale
evolution comes by contracting Eqs.~(\ref{lsab2}) and
(\ref{lEab2}) with $\sigma_{ab}$ and $E_{ab}$ respectively. In
particular we find
\begin{eqnarray}
(\sigma^2)^{.}&=&-{\textstyle{4\over3}}\Theta\sigma^2- {\cal X}-
{\textstyle{1\over2}}H^2\Sigma\,,  \label{lsigma2}\\
(E^2)^{.}&=&-2\Theta E^2- {\textstyle{1\over2}}\mu(1+w){\cal X}-
{\textstyle{1\over2}}\Theta H^2{\cal E}\,, \label{lE2}
\end{eqnarray}
with ${\cal X}=E_{ab}\sigma^{ab}$,
$\Sigma=\sigma_{ab}\eta^a\eta^b$ and ${\cal
E}=E_{ab}\eta^a\eta^b$, while $\eta_a$ is the constant unitary
vector in the direction of the magnetic force lines. The above
system closes with the following propagation equations
\begin{eqnarray}
\dot{{\cal X}}&=&-{\textstyle{5\over3}}\Theta{\cal X}- 2E^2-
{\textstyle{1\over2}}H^2{\cal E}- \mu(1+w)\sigma^2-
{\textstyle{1\over2}}\Theta H^2\Sigma\,,  \label{lcX}\\
\dot{\Sigma}&=&-{\textstyle{2\over3}}\Theta\Sigma- {\cal E}-
{\textstyle{1\over3}}H^2\,,  \label{lSigma}\\ \dot{{\cal
E}}&=&-\Theta{\cal E}- {\textstyle{1\over2}}\mu(1+w)\Sigma-
{\textstyle{1\over3}}\Theta H^2\,,  \label{lcE}
\end{eqnarray}
where the magnetic energy density simply redshifts with the
expansion
\begin{equation}
(H^2)^{.}=-{\textstyle{4\over3}}\Theta H^2\,.  \label{lH2}
\end{equation}

According to Eqs.~(\ref{lsab1}), (\ref{lEab1}) (or equivalently
(\ref{lsab2}), (\ref{lEab2})) the magnetic effects on
$\sigma_{ab}$ and $E_{ab}$ propagate via $\Pi_{ab}=-H_{\langle
a}H_{b\rangle}$. Since both $\sigma_{ab}$ and $E_{ab}$ are also
trace-free tensors, the magnetic effects on $\sigma^2$ and $E^2$
propagate through the last two terms in Eqs.~(\ref{lsigma2}),
(\ref{lE2}), namely via the contractions
$\Sigma=\sigma_{ab}\eta^a\eta^b$ and ${\cal
E}=E_{ab}\eta^a\eta^b$. The latter describe the ``squeezing'' and
the ``stretching'' of the space, along $\eta_a$, that is caused by
the propagating gravitational wave. Crucially, $\eta_a$ is the
direction the magnetic tension acts along. Hence, after the scalar
and the vector modes have been switched off, the only remaining
linear magnetic effect comes from the tension properties of the
field lines.

The scalars $\Sigma$ and ${\cal E}$ are related to spatial
curvature perturbations via the Gauss-Codacci equation.
Contracting Eq.~(\ref{GC}) twice along $\eta_a$ and keeping up to
first order terms we arrive at
\begin{equation}
\Re={\cal E}-{\textstyle{1\over3}}\Theta\Sigma\,,  \label{Re}
\end{equation}
where $\Re$ is the total spatial curvature perturbation along the
magnetic direction.\footnote{In~\cite{MTU} the scalars $\Sigma$
and ${\cal E}$ were related to spatial curvature distortions by
twice contracting the PSTF part of Eq.~(\ref{GC}) along $\eta_a$.
Here we exploit the fact that the magnetic contribution to
3-curvature perturbations (along the direction of the field lines)
vanishes, and relate $\Sigma$ and ${\cal E}$ to the total spatial
curvature perturbation $\Re$. This choice does not alter the
essence of the calculation, but allows for a more transparent
discussion on the subtle role of spatial curvature perturbations.}
Recall that the total contribution of the field to $\Re$ is zero
(see Sec.~3.2), which explains the absence of any magnetic terms
in the above. Also, given that all scalar perturbations have been
switched off and that the background geometry is flat,
Eq.~(\ref{Re}) contains no matter or expansion terms either.

\section{Magnetic effects on gravitational waves}
\subsection{The radiation epoch}
When radiation dominates $w=1/3$, $\mu=3/4t^2$ and $\Theta=3/2t$
(see Eqs.~(\ref{ce}), (\ref{Fe})). Then, according to
Eq.~(\ref{lH2}), the magnetic energy density drops as
\begin{equation}
H^2=H_0^2\left(\frac{t_0}{t}\right)^2\,,  \label{rH2}
\end{equation}
relative to a comoving observer with $u_a=\delta_a^0u_0$. On using
the above result, we find that in the same epoch the energy of the
waves, as they propagate through the magnetized radiation fluid,
is governed by the equations
\begin{eqnarray}
(\sigma^2)'&=&-2t^{-1}\sigma^2- {\cal X}-
{\textstyle{1\over2}}\alpha t^{-2}\Sigma\,,  \label{rlsigma2}\\
(E^2)'&=&-3t^{-1}E^2- {\textstyle{1\over2}}t^{-2}{\cal X}-
{\textstyle{3\over4}}\alpha t^{-3}{\cal E}\,,  \label{rlE2}
\end{eqnarray}
which form a closed system with the set
\begin{eqnarray}
{\cal X}'&=&-{\textstyle{5\over2}}t^{-1}{\cal X}- 2E^2-
t^{-2}\sigma^2- {\textstyle{1\over2}}\alpha t^{-2}{\cal
E}- {\textstyle{3\over4}}\alpha t^{-3}\Sigma\,,  \label{rlcX}\\
\Sigma'&=&-t^{-1}\Sigma- {\cal E}- {\textstyle{1\over3}}\alpha
t^{-2}\,, \label{rlSigma}\\
{\cal E}'&=&-{\textstyle{3\over2}}t^{-1}{\cal E}-
{\textstyle{1\over2}}t^{-2}\Sigma- {\textstyle{1\over2}}\alpha
t^{-3}\,,  \label{rlcE}
\end{eqnarray}
where $\alpha=H_0^2t_0^2$ and the dash indicates derivatives with
respect to coordinate time. Note that the zero suffix corresponds
to the initial time $t_0$. The system
(\ref{rlsigma2})-(\ref{rlcE}) accepts the late-time (i.e.~when
$t\gg t_0$) solutions
\begin{eqnarray}
\sigma^2&=&{\textstyle{1\over9}}\left[\sigma_0^2+ 4E_0^2t_0^2-
2{\cal X}_0t_0- 2\left({\cal E}_0-
{\textstyle{1\over2}}\frac{\Sigma_0}{t_0}-
{\textstyle{1\over6}}H_0^2\right)H_0^2t_0^2\right]\,,
\label{frsigma2}\\ E^2&=&{\textstyle{4\over9}}\left[E_0^2+
{\textstyle{1\over4}}\left(\frac{\sigma_0}{t_0}\right)^2-
{\textstyle{1\over2}}\frac{{\cal X}_0}{t_0}-
{\textstyle{1\over2}}\left({\cal E}_0-
{\textstyle{1\over2}}\frac{\Sigma_0}{t_0}-
{\textstyle{1\over6}}H_0^2\right)H_0^2\right]\left(\frac{t_0}{t}\right)^2\,,
\label{frE2}
\end{eqnarray}
for the magnitudes of $\sigma_{ab}$ and $E_{ab}$ respectively (see
Appendix). Moreover, for radiation the twice contracted
Gauss-Codacci equation (see (\ref{Re})) gives
\begin{equation}
\Re_0={\cal E}_0- {\textstyle{1\over2}}\frac{\Sigma_0}{t_0}\,,
\label{rRe}
\end{equation}
which exactly coincides with the gravitational wave contribution
to the parentheses in Eqs.~(\ref{frsigma2}), (\ref{frE2}). On
using the above, solutions (\ref{frsigma2}), (\ref{frE2})
transform into
\begin{equation}
\sigma^2={\textstyle{1\over9}}\left[\sigma_0^2+ 4E_0^2t_0^2-
2{\cal X}_0t_0\right]-
{\textstyle{2\over9}}\left(\Re_0-{\textstyle{1\over6}}H_0^2\right)H_0^2t_0^2\,,
\label{ltrsigma2}
\end{equation}
and
\begin{equation}
E^2={\textstyle{4\over9}}\left[E_0^2+
{\textstyle{1\over4}}\left(\frac{\sigma_0}{t_0}\right)^2-
{\textstyle{1\over2}}\frac{{\cal
X}_0}{t_0}\right]\left(\frac{t_0}{t}\right)^2-
{\textstyle{2\over9}}\left(\Re_0-{\textstyle{1\over6}}H_0^2\right)H_0^2\left(\frac{t_0}{t}\right)^2\,,
\label{ltrE2}
\end{equation}
respectively. The quantities in brackets describe the
non-magnetized case. One can easily verify this by comparing our
solutions to those of the magnetic-free studies~\cite{DBE,C}

Results (\ref{ltrsigma2}) and (\ref{ltrE2}) show that the field
leaves the evolution rate of $\sigma^2$ and $E^2$ unaffected, but
modifies their magnitudes. Therefore, the overall magnetic impact
depends entirely on the initial conditions and it is sensitive to
the initial value of $\Re$. The latter describes perturbations in
the spatial curvature, along the lines of the field, caused by the
passing gravitational wave. In principle $\Re_0$ can take either
positive or negative values.

\subsection{The dust epoch}
During dust domination $w=0$ and the zero order equations
guarantee that $\Theta=2/t$ and $\mu=4/3t^2$. Accordingly, the
magnetic energy density evolves as
\begin{equation}
H^2=H_0^2\left(\frac{t_0}{t}\right)^{8/3}\,, \label{dH2}
\end{equation}
relative to a comoving observer. The evolution of $\sigma^2$ and
$E^2$ is monitored by the system
\begin{eqnarray}
(\sigma^2)'&=&-{\textstyle{8\over3}}t^{-1}\sigma^2- {\cal X}-
{\textstyle{1\over2}}\beta t^{-8/3}\Sigma\,, \label{dlsigma2}\\
(E^2)'&=&-4t^{-1}E^2- {\textstyle{2\over3}}t^{-2}{\cal X}- \beta
t^{-11/3}{\cal E}\,,  \label{dlE2}
\end{eqnarray}
supplemented by the set
\begin{eqnarray}
{\cal X}'&=&-{\textstyle{10\over3}}t^{-1}{\cal X}- 2E^2-
{\textstyle{4\over3}}t^{-2}\sigma^2- {\textstyle{1\over2}}\beta
t^{-8/3}{\cal E}- \beta t^{-11/3}\Sigma\,,  \label{rdlcX}\\
\Sigma'&=&-{\textstyle{4\over3}}t^{-1}\Sigma- {\cal E}-
{\textstyle{1\over3}}\beta t^{-8/3}\,,
\label{dlSigma}\\
{\cal E}'&=&-2t^{-1}{\cal E}- {\textstyle{2\over3}}t^{-2}\Sigma-
{\textstyle{2\over3}}\beta t^{-11/3}\,, \label{dlcE}
\end{eqnarray}
with $\beta=H_0^2t_0^{8/3}$. Similarly to the radiation case
before, the above has the following late-time solution
\begin{eqnarray}
\sigma^2&=&{\textstyle{4\over25}}\left[\sigma_0^2+
{\textstyle{9\over4}}E_0^2t_0^2- {\textstyle{3\over2}}{\cal
X}_0t_0\right]\left(\frac{t_0}{t}\right)^{2/3}-
{\textstyle{9\over50}}\left({\cal
E}_0-{\textstyle{2\over3}}\frac{\Sigma_0}{t_0}
-{\textstyle{1\over6}}H_0^2\right)H_0^2t_0^2\left(\frac{t_0}{t}\right)^{2/3}\,,
\label{fdsigma2}\\ E^2&=&{\textstyle{9\over25}}\left[E_0^2+
{\textstyle{4\over9}}\left(\frac{\sigma_0}{t_0}\right)^2-
{\textstyle{2\over3}}\frac{{\cal
X}}{t_0}\right]\left(\frac{t_0}{t}\right)^{8/3}-
{\textstyle{9\over50}}\left({\cal
E}_0-{\textstyle{2\over3}}\frac{\Sigma_0}{t_0}
-{\textstyle{1\over6}}H_0^2\right)H_0^2\left(\frac{t_0}{t}\right)^{8/3}\,.
\label{fdE2}
\end{eqnarray}
In the dust era the twice contracted Gauss-Codacci equation
(\ref{Re}) gives
\begin{equation}
\Re_0={\cal E}_0- {\textstyle{2\over3}}\frac{\Sigma_0}{t_0}\,.
\label{dRe}
\end{equation}
which again coincides with the wave contribution to the
parentheses in Eqs.~(\ref{fdsigma2}), (\ref{fdE2}). On using the
above we may recast Eqs.~(\ref{fdsigma2}) and (\ref{fdE2}) into
\begin{equation}
\sigma^2={\textstyle{4\over25}}\left[\sigma_0^2+
{\textstyle{9\over4}}E_0^2t_0^2- {\textstyle{3\over2}}{\cal
X}_0t_0\right]\left(\frac{t_0}{t}\right)^{2/3}-
{\textstyle{9\over50}}\left(\Re_0
-{\textstyle{1\over6}}H_0^2\right)H_0^2t_0^2\left(\frac{t_0}{t}\right)^{2/3}\,,
\label{ltdsigma2}
\end{equation}
and
\begin{equation}
E^2={\textstyle{9\over25}}\left[E_0^2+
{\textstyle{4\over9}}\left(\frac{\sigma_0}{t_0}\right)^2-
{\textstyle{2\over3}}\frac{{\cal
X}}{t_0}\right]\left(\frac{t_0}{t}\right)^{8/3}-
{\textstyle{9\over50}}\left(\Re_0
-{\textstyle{1\over6}}H_0^2\right)H_0^2\left(\frac{t_0}{t}\right)^{8/3}\,,
\label{ltdE2}
\end{equation}
respectively, where again the brackets describe the non-magnetized
case~\cite{DBE,C}. As with radiation before, the overall magnetic
effect depends entirely on the initial conditions. Note the
magneto-curvature terms in Eqs.~(\ref{ltrsigma2}), (\ref{ltrE2})
and (\ref{ltdsigma2}), (\ref{ltdE2}). Qualitatively, this
magneto-geometrical effect tends to reduce the energy of the wave
when the initial curvature distortion is positive (i.e.~for
$\Re_0>0$), but increases both $\sigma^2$ and $E^2$ if $\Re_0<0$.
Quantitatively, the effect gets stronger with increasing curvature
distortion. We will return to this non-trivial behavior of the
field later.

\subsection{Production of gravitational waves}
Consider the dust dominated era and assume that there are no
gravitational waves originally present. In this case the initial
conditions are $\sigma_0^2=E_0^2={\cal X}_0={\cal
E}_0=\Sigma_0=0$. Then solutions (\ref{fdsigma2}) and (\ref{fdE2})
give
\begin{eqnarray}
\sigma^2&=&{\textstyle{3\over10^2}}H_0^4t_0^2\left(\frac{t_0}{t}\right)^{2/3}\,,
\label{prsigma2}\\
E^2&=&{\textstyle{3\over10^2}}H_0^4\left(\frac{t_0}{t}\right)^{8/3}\,,
\label{prE2}
\end{eqnarray}
with an analogous result for dust. Thus, the magnetic presence has
led to gravity wave perturbations. Note that this magnetically
induced production of gravitational waves results purely from the
anisotropic nature of the field. Qualitatively, the tension
properties of the magnetic field lines are of no consequence. That
is to say any other anisotropic source would have also triggered
gravitational wave distortions. Indeed, results (\ref{prsigma2}),
(\ref{prE2}) remain unchanged when the minus sign of the magnetic
terms in Eqs.~(\ref{lsigma2})-(\ref{lcE}) is replaced by a plus,
namely when we switch from tension to ordinary positive pressure.
Assuming the presence of a cosmological magnetic field at
recombination we may use result (\ref{prsigma2}) to estimate the
strength of the induced gravitational wave. Adopting an upper
limit of $10^{-9}~G$, in today's values (see~\cite{BFS}), we find
an induced wave of the order of $10^{-70}~GeV^4$ also in today's
values. The latter lies twelve orders of magnitude bellow the
energy density of the inducing magnetic field.

\subsection{Boosting and damping of gravitational waves}
Let us now return to the magneto-curvature terms in
Eqs.~(\ref{ltrsigma2}), (\ref{ltrE2}) and (\ref{ltdsigma2}),
(\ref{ltdE2}). Their presence allows for the possibility of a zero
overall magnetic effect if $\Re=H^2/6$ initially, namely if the
curvature distortion along the field lines ``equals'' the
isotropic magnetic pressure. Alone, the magneto-curvature terms
reduce the energy of the wave if $\Re_0>0$, but lead to an
increase when $\Re_0<0$. The effect results directly from the
tension properties of the magnetic force lines. In fact, if
tension were replaced by ordinary pressure in
Eqs.~(\ref{lsigma2})-(\ref{lcE}) the effect is reversed. This
intricate behaviour can be seen as the field's reaction to spatial
curvature deformations. More specifically, the non-linear
Gauss-Codacci equation (\ref{GC}) gives
\begin{equation}
\Re={\cal R}_{ab}\eta^a\eta^b={\textstyle{2\over3}}\sigma^2-
{\textstyle{1\over3}}\Theta\sigma_{ab}\eta^a\eta^b+
\sigma_{c\langle a}\sigma^c{}_{b\rangle}\eta^a\eta^b+
E_{ab}\eta^a\eta^b\,,  \label{cGC2}
\end{equation}
for the total perturbation in the spatial curvature along the
direction of the field lines. Note the absence of any magnetic
terms in the above since the contribution of the field to ${\cal
R}_{ab}\eta^a\eta^b$ is zero (see Sec.~3.2). Also, given that both
matter and volume expansion perturbations have been switched off
(see Sec.~3.3), $2\mu-\Theta^2/3=0$ due to the background
flatness. The last three terms in the right-hand side of
Eq.~(\ref{cGC2}) describe the ``stretching'' and the ``squeezing''
of the space (in the direction of the field lines) caused by the
passing gravitational wave and take positive or negative values.
However on average, say over one oscillation period, their
contribution to ${\cal R}_{ab}\eta^a\eta^b$ amounts to zero,
leaving $\sigma^2$ as the only wave input to curvature
deformations. Thus, by increasing $\sigma^2$ when the initial
curvature is negative and by decreasing it when $\Re_0>0$, the
field tends to minimise the curvature perturbation along its own
direction. This magnetic reaction to curvature distortions is
indicative of the tension properties of the field lines, that is
of their tendency to remain as straight as possible. Analogous
magneto-curvature effects have also been identified on the
expansion of magnetised cosmologies (see~\cite{MT}) and might be
interpreted as an indication of a magnetic preference for spatial
flatness~\cite{T,T1}.

\section{Discussion}
The question of how magnetic fields interact with gravitational
radiation is as timely as ever in view of the forthcoming gravity
wave detection experiments and the ubiquity of magnetic fields in
the universe. The lack of extensive research on the subject and
the unique features of magnetic fields make the outcome of any
such study difficult to foresee, while it could also probe unknown
as yet aspects of fundamental physics. In the present article we
have considered the impact of a large-scale magnetic field on
gravity waves in the cosmological context. Our starting point was
a spatially flat FRW background which was subsequently perturbed
by weak gravitational waves and a weak magnetic field. We adopted
a geometrical approach and employed the covariant formalism to
examine the field effects on $\sigma^2$ and $E^2$, the scalars
that directly describe the energy density of gravitational
radiation. Throughout the analysis we maintained the pure tensor
nature of the perturbed variables, employed scalars that were
invariantly constructed from these tensors, and did not assume any
a priori relation between the magnetic and the gravitational wave
anisotropies. The geometrical nature of our approach brought to
the fore the tension properties of magnetic fields and revealed
their subtle interconnection with the spatial geometry of the
magnetised spacetime. We found that the overall impact of the
field depends on the initial set up and is particularly sensitive
to the initial curvature deformation as measured along the
direction of the field. In the absence of gravitational waves the
magnetic presence led to wave production. Given the generically
anisotropic nature of magnetic fields this is not a surprising
result. The non-trivial effects came from the intricate coupling
between the geometry and the tension properties of the field. The
presence of the field was found to suppress the energy of
gravitational waves when the initial curvature perturbation was
positive, but led to a boost in the case of positive curvature
deformation. Overall, the field seemed to react to the curvature
distortion caused by the propagating wave and tried to keep it
down to a minimum by modulating the wave's energy accordingly.

The complete dependence of the magnetic effects on the initial
conditions is rather unfortunate, as at this stage it is not clear
what are the best physically motivated initial configurations. The
ambiguity stems from the fact that, to linear order, $\Re_0$
depends on $\Sigma_0$ and ${\cal E}_0$ (see Eqs.~(\ref{rRe}),
(\ref{dRe})) which can take either positive or negative values. Ii
is likely that a non-linear analysis would incorporate
$\sigma^2_0$, namely the wave's original energy, thus favouring
one initial configuration at the expense of the other. That aside,
some interesting questions emerge when one is allowed to speculate
on the basis of the linear results. Astrophysical magnetic fields,
for example, are quite widespread and also considerably strong.
Typical spiral galaxies carry extensive fields of few $\mu G$ and
compact stars can locally support magnetic fields as strong as
$10^{16}\,G$. If the field presence were to boost gravity wave
perturbations in general, then, depending on the efficiency of the
mechanism of course, one might think that detecting gravitational
waves should have been a rather straightforward task. If, on the
other hand, magnetic fields can suppress gravity waves, the
ubiquity of cosmic magnetism could prove a considerable setback
for the forthcoming gravity wave detection projects.

\section*{Acknowledgements}
The author wishes to thank Peter Dunsby, George Ellis and Roy
Maartens for helpful discussions and comments. This work was
supported by a Sida/NRF grant.

\section*{Appendix: Solutions}
One obtains the solutions (\ref{frsigma2}), (\ref{frE2}) and
(\ref{fdsigma2}), (\ref{fdE2}) after a straightforward but rather
tedious calculation. Here, we guide the interested reader through
the intermediate steps. To begin with, Eqs.~(\ref{rlsigma2}),
(\ref{rlE2}) and the set (\ref{rlcX})-(\ref{rlcE}) accept the
solutions
\begin{eqnarray}
\sigma^2&=&-{\textstyle{1\over2}}{\cal C}_1+ {\cal C}_2t^{-3}-
2{\cal C}_3t^{-3/2}- {\textstyle{4\over3}}{\cal C}_4t^{-1}+
{\textstyle{4\over3}}{\cal C}_5t^{-5/2}+
{\textstyle{4\over3}}\alpha^2t^{-2}\,, \label{rsigma2}\\
E^2&=&-{\textstyle{1\over2}}{\cal C}_1t^{-2}+
{\textstyle{1\over4}}{\cal C}_2t^{-5}+ {\cal C}_3t^{-7/2}+
{\textstyle{1\over3}}{\cal C}_4t^{-3}+ {\textstyle{1\over6}}{\cal
C}_5t^{-9/2}+ {\textstyle{1\over12}}\alpha^2t^{-4}\,, \label{rE2}
\end{eqnarray}
and
\begin{eqnarray}
{\cal X}&=&{\cal C}_1t^{-1}+ {\cal C}_2t^{-4}+ {\cal C}_3t^{-5/2}+
{\cal C}_4t^{-2}+ {\cal C}_5t^{-7/2}+
{\textstyle{2\over3}}\alpha^2t^{-3}\,. \label{rcX}\\
\Sigma&=&{\textstyle{2\over3}}\alpha^{-1}{\cal C}_4-
{\textstyle{2\over3}}\alpha^{-1}{\cal C}_5t^{-3/2}-
{\textstyle{4\over3}}\alpha^{-1}t^{-1}\,, \label{rSigma}\\ {\cal
E}&=&-{\textstyle{2\over3}}\alpha^{-1}{\cal C}_4t^{-1}-
{\textstyle{1\over3}}\alpha {\cal C}_5t^{-5/2}-
{\textstyle{1\over3}}\alpha t^{-2}\,,  \label{rcE}
\end{eqnarray}
respectively, where $\alpha=H_0^2t_0^2$ and ${\cal C}_\imath$
(with $\imath=1\,,\ldots,5$) are the integration constants.
Solutions (\ref{rcE}), (\ref{rSigma}) immediately provide the
expressions
\begin{eqnarray}
{\cal C}_4&=&-{\textstyle{1\over2}}\left(2{\cal E}_0-
\Sigma_0t_0^{-1}-{\textstyle{2\over3}}H_0^2\right)H_0^2t_0^3\,,
\label{C4}\\ {\cal C}_5&=&-\left({\cal E}_0+
\Sigma_0t_0^{-1}+{\textstyle{5\over3}}H_0^2\right)H_0^2t_0^{9/2}\,.
\label{C5}
\end{eqnarray}
which substituted into Eqs.~(\ref{rcX})-(\ref{rcE}) lead, after a
lengthy but straightforward calculation, to the rest of the
integration constants
\begin{eqnarray}
{\cal C}_1&=&-{\textstyle{4\over9}}\left[2E_0^2t_0^2+
{\textstyle{1\over2}}\sigma_0^2- {\cal X}_0t_0- \left({\cal E}_0-
{\textstyle{1\over2}}\Sigma_0t_0^{-1}-
{\textstyle{1\over6}}H_0^2\right)H_0^2t_0^2\right]\,,  \label{C1}\\
{\cal C}_2&=&{\textstyle{4\over9}}\left[E_0^2t_0^2+ \sigma_0^2+
{\cal X}_0t_0+ \left({\textstyle{5\over2}}{\cal E}_0+
{\textstyle{5\over2}}\Sigma_0t_0^{-1}+
{\textstyle{25\over12}}H_0^2\right)H_0^2t_0^2\right]t_0^3\,,
\label{C2}\\ {\cal C}_3&=&{\textstyle{1\over9}}\left[4E_0^2t_0^2-
2\sigma_0^2+ {\cal X}_0t_0+ \left(4{\cal E}_0-
{\textstyle{7\over2}}\Sigma_0t_0^{-1}-
{\textstyle{5\over3}}H_0^2\right)H_0^2t_0^2\right]t_0^{3/2}\,.
\label{C3}
\end{eqnarray}
On using results (\ref{C4})-(\ref{C3}), we arrive at the late-time
(i.e.~for $t\gg t_0$) solutions (\ref{frsigma2}) and (\ref{frE2})
respectively.

For the dust era we start from the system
(\ref{dlsigma2})-(\ref{dlcE}) and then proceed in a completely
analogous way to obtain the late-time solutions (\ref{ltdsigma2})
and (\ref{ltdE2}). Here, we only provide the full solutions to the
set (\ref{dlsigma2})-(\ref{dlcE}). They respectively are
\begin{eqnarray}
\sigma^2&=&{\textstyle{3\over4}}\mathrm{C}_1t^{-4}-
{\textstyle{1\over2}}\mathrm{C}_2t^{-2/3}- 3\mathrm{C}_3t^{-7/3}-
2\mathrm{C}_4t^{-2}+ {\textstyle{6\over7}}\mathrm{C}_5t^{-11/3}+
3\beta^2t^{-10/3}\,, \label{dsigma2}\\
E^2&=&{\textstyle{1\over3}}\mathrm{C}_1t^{-6}-
{\textstyle{1\over2}}\mathrm{C}_2t^{-8/3}+ 2\mathrm{C}_3t^{-13/3}+
\mathrm{C}_4t^{-4}+ {\textstyle{2\over7}}\mathrm{C}_5t^{-17/3}+
{\textstyle{3\over4}}\beta^2t^{-16/3}\,,  \label{dE2}\\
{\cal X}&=&\mathrm{C}_1t^{-5}+ \mathrm{C}_2t^{-5/3}+
\mathrm{C}_3t^{-10/3}+ \mathrm{C}_4t^{-3}+ \mathrm{C}_5t^{-14/3}+
3\beta^2t^{-13/3}\,, \label{dcX}\\
\Sigma&=&{\textstyle{2\over3}}\beta^{-1}\mathrm{C}_4t^{-1/3}-
{\textstyle{2\over7}}\beta^{-1}\mathrm{C}_5t^{-2}- 2\beta
t^{-5/3}\,, \label{dSigma}\\ {\cal
E}&=&-{\textstyle{2\over3}}\beta^{-1}\mathrm{C}_4t^{-4/3}-
{\textstyle{4\over21}}\beta^{-1}\mathrm{C}_5t^{-3}- 2\beta
t^{-8/3}\,, \label{dcE}
\end{eqnarray}
where $\beta=H_0^2t_0^{8/3}$ and $\mathrm{C}_i$ (with
$i=1\,,\ldots,5$) are the integration constants.

\end{document}